\documentclass[10pt, preprint, pdftex]{emulateapj}	

\usepackage{color}
\usepackage{amsmath}	

\usepackage{style_wei_symbols}
\usepackage{style_wei_2018-05-05_Fig}
\usepackage{style_wei_journal_names}

\newcommand*{\st}{space\,--\,time}

\usepackage{natbib}
\def\editor#1{{#1}}
\def\revise#1{{#1}}

\begin{document}


\title{A Truly Global \editor{Extreme Ultraviolet} 
Wave from the SOL2017-09-10 X8.2\editor{+} Solar Flare-\editor{Coronal Mass Ejection}}





\author{Wei Liu\altaffilmark{1,2,3},		
Meng Jin\altaffilmark{1,4},	
Cooper Downs\altaffilmark{5}, Leon Ofman\altaffilmark{6,7,8},	
Mark \editor{C.~M.} Cheung\altaffilmark{1}, and Nariaki \editor{V.~}Nitta\altaffilmark{1}
}

\altaffiltext{1}{Lockheed Martin Solar and Astrophysics Laboratory, Bldg.~252,  
  3251 Hanover Street, Palo Alto, CA 94304, USA; weiliu@lmsal.com}
\altaffiltext{2}{Bay Area Environmental Research Institute, NASA Research Park, Mailstop 18-4, Moffett Field, CA 94035-0001, USA}
\altaffiltext{3}{W.~W.~Hansen Experimental Physics Laboratory, Stanford University, Stanford, CA 94305, USA}
\altaffiltext{4}{SETI Institute, 189 N Bernardo Ave suite 200, Mountain View, CA 94043, USA}
\altaffiltext{5}{Predictive Science Inc., 9990 Mesa Rim Road, Suite 170, San Diego, CA 92121, USA}
\altaffiltext{6}{Catholic University of America, Washington, DC 20064, USA} 
\altaffiltext{7}{NASA Goddard Space Flight Center, Code 671, Greenbelt, MD 20771, USA} 
\altaffiltext{8}{Visiting, Department of Geosciences, Tel Aviv University, Tel Aviv, Israel}

\shorttitle{Truly Global EUV Wave}
\shortauthors{Liu et al.}
\slugcomment{Accepted by ApJ Letters, July 24, 2018}

\begin{abstract}	

We report \sdo/AIA observations of an extraordinary global extreme ultraviolet (EUV) wave
triggered by the X8.2\editor{+} flare-CME eruption on 2017 September 10.	
This was one of the best EUV waves ever observed with modern instruments,		
yet \editor{it was} likely the last one of such magnitudes of Solar Cycle~24 as the Sun heads toward the minimum. 
Its remarkable characteristics include \editor{the following.} 
(1) The wave was observed, for the first time, to traverse the full-Sun corona 
over the entire visible solar disk and off-limb circumference, 
manifesting a truly global nature, 	
owing to its exceptionally large	
amplitude,	
e.g., with EUV enhancements by up to 300\% 	
at $1.1\Rsun$ from the eruption. 	
(2) 
This leads to strong transmissions (\editor{in addition to} 
commonly observed reflections)
in and out of both polar coronal holes, which are usually devoid of EUV waves.
It has elevated wave speeds $>$$2000\kmps$ within them,
consistent with the expected higher fast-mode magnetosonic \editor{wave} speeds.
The coronal holes essentially serve as new ``radiation centers"
for the waves being refracted out of them, which then	
travel toward the equator and collide head-on, causing additional EUV enhancements. 	
(3) The wave produces significant compressional	
heating to local plasma
upon its impact,	
indicated by long-lasting EUV intensity changes and differential emission measure increases 
at higher temperatures (e.g., $\log T=6.2$) accompanied by decreases at lower temperatures (\editor{e.g., }$\log T=6.0$). 
These characteristics signify the potential of such EUV waves for novel magnetic and thermal diagnostics 
of the solar corona {\it on global scales}. 
\end{abstract}


\section{Introduction}
\label{sect_intro}


Global extreme ultraviolet (EUV) waves 	
\citep{ThompsonB.EIT-wave-discover.1998GeoRL..25.2465T}
are intensity disturbances expanding across a sizable 
fraction of the solar corona. 	
They are one of the most spectacular manifestations of solar eruptions
involving flares and coronal mass ejections (CMEs).
After a decade-long 	
debate with extensive observations
\citep[e.g.,][]{Patsourakos.EUVI-fast-mode-wave.2009ApJ...700L.182P, 	
LiuW.AIA-1st-EITwave.2010ApJ...723L..53L, 
TemmerM.3D.stereo.EUVwv.2011SoPh..273..421T,
NittaN.AIA.wave.stat.2013ApJ...776...58N, 
MuhrN.STEREO.EUVwave.stat.2014SoPh..289.4563M, LongDave.EITwv.stat.CorPITA.2017SoPh..292..185L}
and modeling 
\citep{Wu.EIT-fast-MHD-wave.2001JGR...10625089W,
ChenPF.EIT-wave-MHD.2002ApJ...572L..99C,
OfmanThompson.EIT-wave-fast-mode.2002ApJ...574..440O, Ofman.wave.2007ApJ...655.1134O, 	
DownsC.MHD.EUVIwave.2011ApJ...728....2D, DownsC.MHD.2010-06-13-AIA-wave.2012ApJ...750..134D},
it was recognized that their physical nature	
embodies	
a	
fast-mode magnetosonic wave component, often accompanied or driven by another	
non-wave component due to CME-caused reconfiguration 
\citep[see reviews by][]{
Wills-Davey.EIT-wave-review.2009SSRv..149..325W,
GallagherP.LongD.EIT.wave.review.2011SSRv..158..365G,
ZhukovA.EIT.wave.review.STEREO.2011JASTP..73.1096Z, 	
Patsourakos.Vourlidas.EIT-wave-review.2012SoPh..281..187P,	
LiuW.OfmanL.EUV.wave.review.2014SoPh..289.3233L, 
WarmuthA.gloabl.coronal.wave.review.2015LRSP...12....3W,
ChenPengfei.EITwv.review.AGU.Champman.Conf.2016GMS...216..381C,
LongDave.review.EUVwv.nature.2017SoPh..292....7L}.

We report here \editor{Solar Dynamics Observatory/Atmospheric Imaging Assembly} (\sdo/AIA)
observations of an extraordinary EUV wave 	
on 2017 September 10.
This was the first detection of a {\it truly global} EUV wave that,
with its exceptionally large amplitude, covered
the full-Sun corona, 	
including traveling in and out of both polar coronal holes (CHs). 	
It produced significant thermal perturbations to the coronal plasma lasting for hours. 	
These characteristics 	
allow waves	of such extreme magnitudes to serve as 	
probes to diagnose the solar corona on global scales, 
a yet under-explored subject \citep{WestMatthew.EIT-wave-seismology.2011ApJ...730..122W,
KwonR_EUV.wave.seismology.2013ApJ...776...55K,
LongDave.EITwv.seismology.trans-eq-loop.2017AA...603A.101L}. 	

	%


\section{Observational Overview}	
\label{sect_overview}


%
 \begin{figure*}[thbp]      
 \begin{center}
 \includegraphics[width=0.95\textwidth]{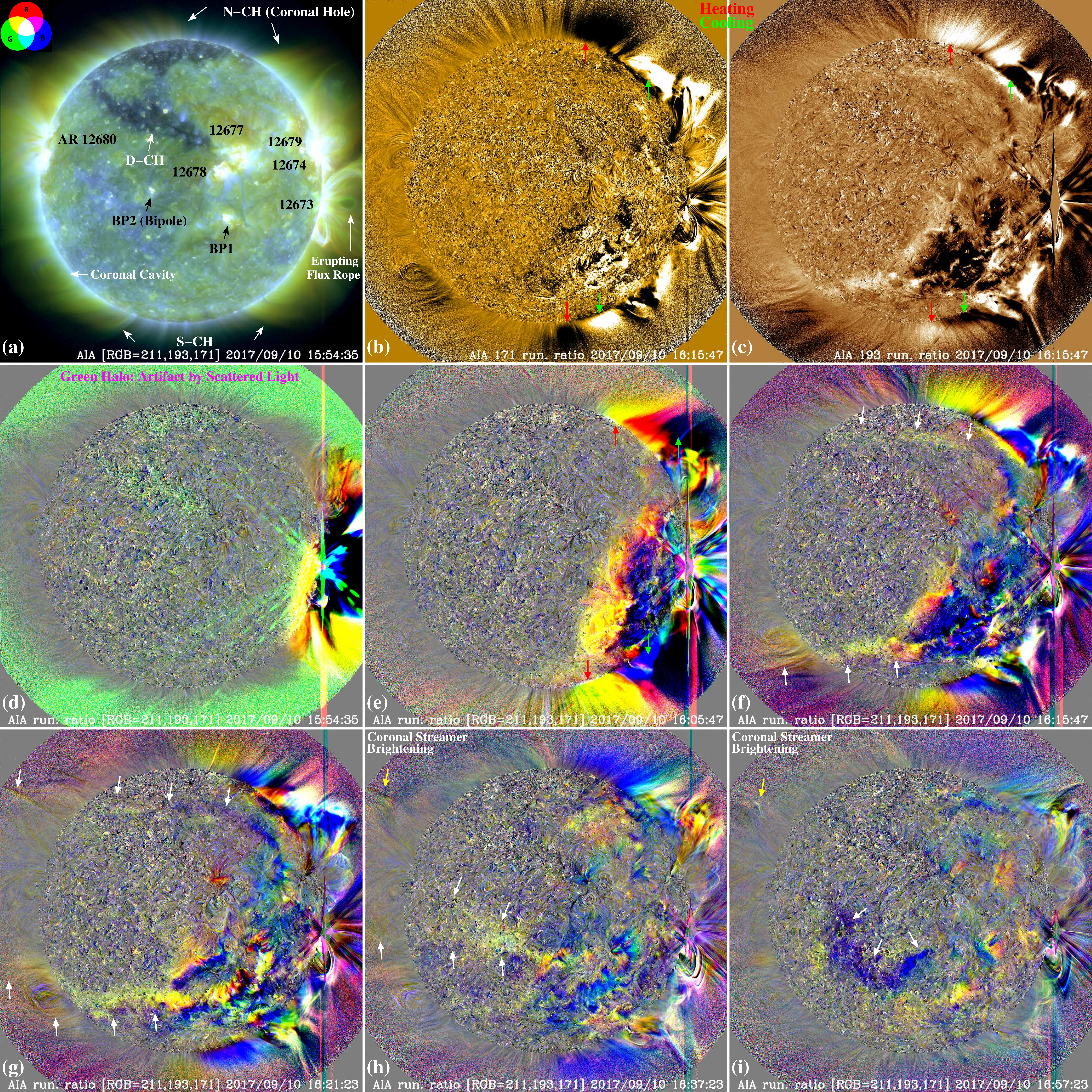} 
 \end{center}
 \caption[]{
 \sdo/AIA overview of the 2017 September 10 global EUV wave.
  (a)~Composite tri-color image, 211~\AA\ (red), 193~\AA\ (green), 171~\AA\ (blue),	
 showing the erupting flux-rope and	 
 notable structures, including ARs, North-/South- polar and on-Disk Coronal Holes (N-/S-, D-CH),
 bipolar regions (BP1, BP2),	
 and a coronal cavity.
  (b)--(c)~Simultaneous 171 and 193~\AA\ running-ratio images showing the global wave. 
  (d)--(i)	
 Running-ratio tri-color	
 images showing its evolution,	
 with plasma heating (cooling) indicated in yellow or red (blue).	
 Short white arrows in (f)--(i) mark wave fronts	
 reflected/refracted from the polar CHs. 
 \editor{An animation of this figure is provided in the online Journal.}
 } \label{mosaic.eps}
 \end{figure*}
%
The event of interest (SOL2017-09-10T16:06) occurred 	
in Active Region (AR) 12673 at the west limb, associated with an X8.2\editor{+} flare and a fast CME, 
which themselves were extremely remarkable 	
in many respects, including being the second largest flare
and causing the second ground-level enhancement 	
event of Solar Cycle 24
\revise{\citep[e.g.,][]{
ChenBin.EOVSA.1st.result.2017Sep10X8.2018arXiv180702498G,
GopalswamyNat.2017Sep10X8.GLE.SEPs.2018arXiv180709906G,
GuoJingNan.2017Sep10.CME.SEP.Mars.2018arXiv180300461G,
KurtViktoria.2017Sep10X8.GLE.2018arXiv180600226K,
LiYing.2017Sep10X8.current.sheet.EIS.AIA.2018ApJ...853L..15L,
OmodeiNicola.2017Sep10-X8.Fermi.2018arXiv180307654O,
PolitoVanessa.2017Sep10X8.EIS.Fe24.non-gaussian.2018arXiv180709361P,
WarrenHarry.2017Sep10X8.CS.EIS.2018ApJ...854..122W, 
YanXiaoLi.2017Sep10X8.flux.rope.reconn.2018ApJ...853L..18Y}.}
	%

The flare occurred at 15:35~UT and peaked at 16:06~UT  
(\Fig{vcut_slice.eps}(b)).		
The impulsive phase (15:52\,--\,16:06~UT) started cotemporally	
with 	
the onset of the rapid ascent \editor{(at an acceleration of $9.5 \pm 0.7 \kmpss$;
cf., \citealt{Veronig.2017Sep10X8.euv.wave.generation.2018ApJ})}
and lateral expansion of the teardrop-shaped flux-rope, 
giving birth to the CME 	
and global EUV wave, which was
observed by \sdo/AIA, \goes-16/SUVI 
(\citealt{SeatonDan.2017Sep10.X8.SUVI.high.corona.2018ApJ...852L...9S}; \editor{T. Podladchikova et al., in preparation}),
and PROBA-2/SWAP	
\citep{GoryaevF.AIA.brighten-darkening.2017Sep6-10.flares.2018ApJ...856L..38G}.
\editor{Its radio signatures were detected by the Low-Frequency Array \citep[LOFAR;][]{Morosan.LOFAR.2017Sep10X8.2018NatAstron}.}

\Fig{mosaic.eps} and its online movies give an overview of the EUV wave evolution.
By 15:52~UT, 	
it is clearly visible surrounding	
the eruption 	
and starts to propagate in all directions. {\it Off-limb}, the wave travels both north- and southward,
traverses both polar CHs and a coronal cavity on the southeast limb, 
finally arrives at AR~12680 on the east limb to the opposite end of the solar diameter from the eruption, 
covering the entire 	
solar limb circumference. 
{\it On-disk}, the wave front begins with a circular shape,
evolves into a reverse-C	
shape comprising of the eastward primary wave
and equatorward secondary waves reflected and refracted from the polar CHs,
which eventually collide at low latitudes.

There are a variety of smaller-scale secondary waves due to reflections and/or refractions, 
wherever the primary or secondary wave encounters structures of differing magnetic-field strengths
and thus fast-mode magnetosonic speeds.	
There are also \revise{ubiquitous} stationary brightenings, 
dimmings \revise{\citep{ZhukovA.global.CME.2007ApJ...664L.131Z}}, 
and oscillations, some lasting for more than two hours,
e.g., at the coronal streamer 	
\revise{and CH boundaries (\Figs{mosaic.eps}--\ref{limb_slice.eps})}.
Collectively, every corner of the full-Sun corona is essentially traversed 	
by this wave.
Its thermal effect 	
is manifested in anti-correlated, large intensity variations between warm (e.g., 193/211~\AA) and cool (171~\AA) channels,
indicative of adiabatic heating and cooling by wave compression and rarefaction.

\section{Global Wave Characteristics}
\label{sect_wave-prop}

\subsection{Off-limb Wave Propagation}
\label{sect_off-limb}

 \begin{figure*}[thbp]      
 \begin{center}
 \epsscale{0.3}
 \includegraphics[height=2.5in]{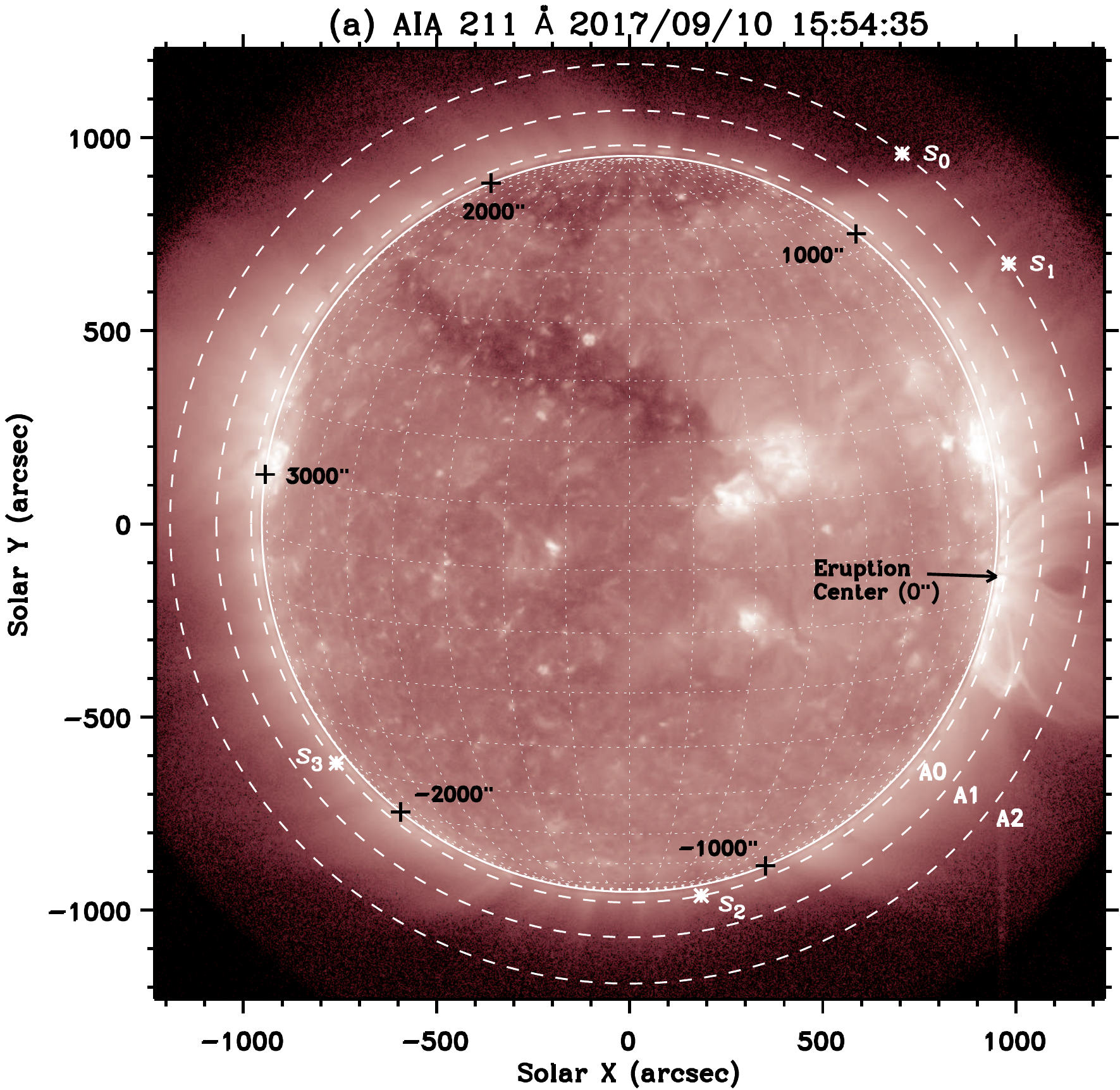} 
 \epsscale{1.}		
 \\
 \includegraphics[width=0.85\textwidth]{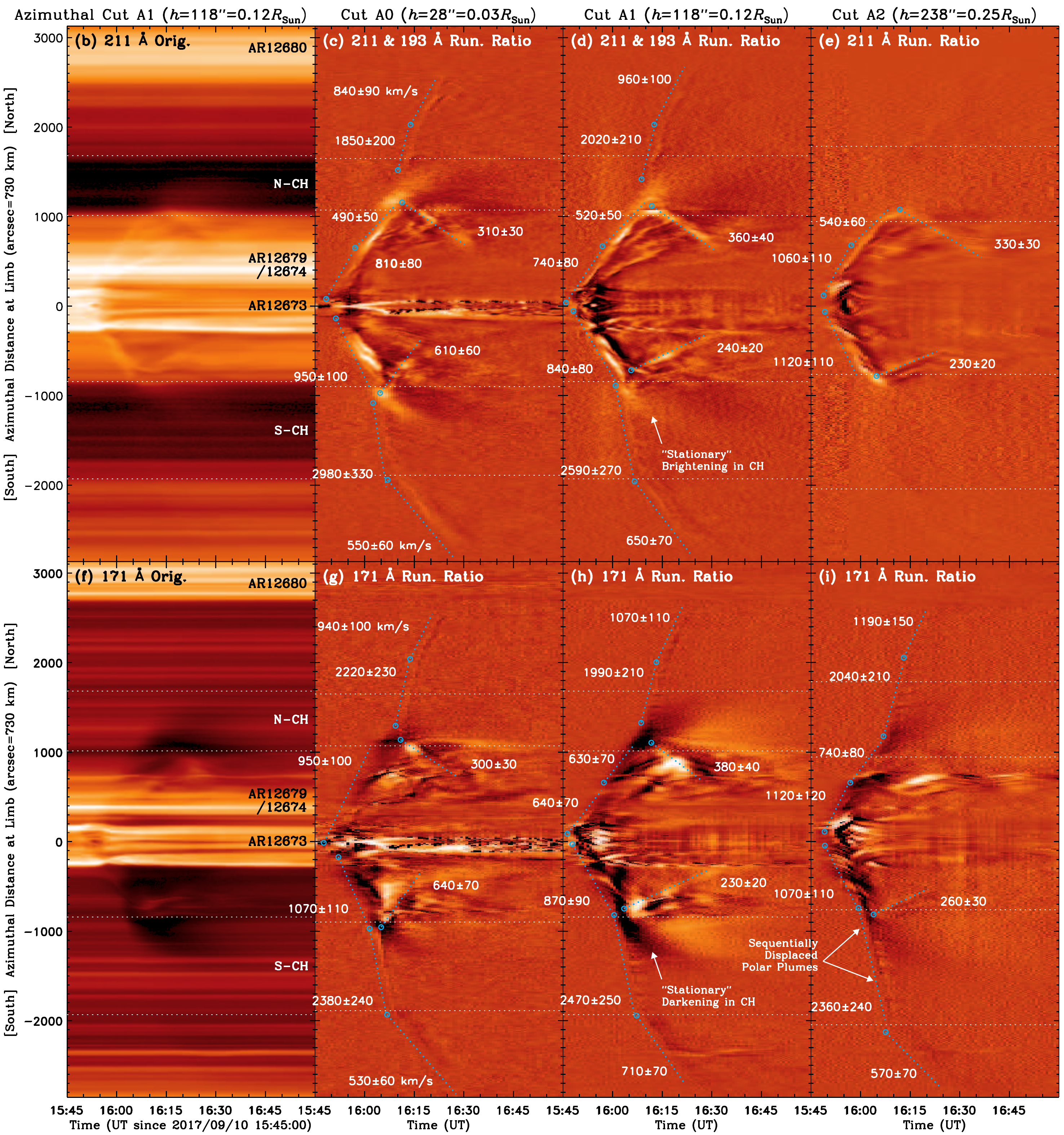} 
 \end{center}
 \vspace{-0.15in}
 \caption[]{
 Off-limb wave propagation.
 (a) Context AIA 211~\AA\ image. 	
 (b)--(i) Space\,--\,time plots from the three 	
 azimuthal cuts shown in (a) 	
 in original intensity (left) and running-ratio (middle/right)	
 at 211/193~\AA\ (top) and 171~\AA\ (bottom).
 The horizontal dotted lines delineate the polar CH boundaries.
 In (c) and (d), 211 and 193~\AA\ data are shown, respectively, between and beyond the two innermost horizontal lines near $\pm 1000\arcsec$.	
 Unless otherwise noted, cyan-color dotted lines starting with an open circle in all \st\ plots of this article 
 are 	
 linear fits to the wave front positions, 	
 labeled with fitted speeds in $\kmps$,
 and are shifted back in time by $-100\s$ here 
 and by $-500\s$ in \Figs{vcut_slice.eps}--\ref{sector_slice.eps} to avoid obscuring the original data.	
 } \label{limb_slice.eps}
 \end{figure*}
%
To track the off-limb wave propagation,	
we selected three azimuthal cuts 
\citep[A0--A2;	
cf.,][]{LiuW.cavity-oscil.2012ApJ...753...52L} 
at constant heights above the limb over the entire global circumference.
\Fig{limb_slice.eps} shows the resulting \st\ diagrams, 
where the distance along the cut is radially mapped onto the limb and measured 
in the counter-clockwise direction from 
the eruption center. 	
The left panels show original intensities,	
where bright ARs and (dark) CHs are identified.
The rest of the panels show running-ratios,	
where the EUV wave front is evident, 	
generally as brightening
at 193/211~\AA\ and darkening at 171~\AA. 	
Upon the wave arrival at the North-/South- polar CH (N-/S-CH) boundary,
both reflection and transmission occur, together with
long-lasting stationary brightening or darkening	
followed by recovery in an opposite direction.	

We tracked 	
traveling intensity changes 
in \st\
and applied piecewise linear fits (using the original cut distance)
to obtain the wave speeds.
The speeds measured at 171, 193, and 211~\AA\ generally agree within uncertainties.
They typically begin with	
700\,--\,$1100\kmps$ near	
the eruption, increase to 1800\,--\,$2600\kmps$ 	
within the polar CHs, 	
and drop back to 500\,--\,$1100\kmps$ upon exiting (or slightly beyond) them.	
The reflection speeds are 300\,--\,$400\kmps$ at the N-CH
and 200\,--\,$600\kmps$ at the S-CH, both being fractionally slower, 
by $20\%-70\%$,	
than the primary wave	
\citep[cf.,][]{OlmedoOscar.2011Feb15_X2.AIA.EUV.wave.2012ApJ...756..143O}.
This is likely because 
(i) the primary wave is shocked, 
while the reflection is a quasi-linear wave at the 	
local fast-mode speed, 	
and/or (ii) the reflection propagates in a different direction and thus a different plane-of-sky (POS) projected speed.

\begin{figure}[thbp]      
 \begin{center}
 \includegraphics[height=2.3in]{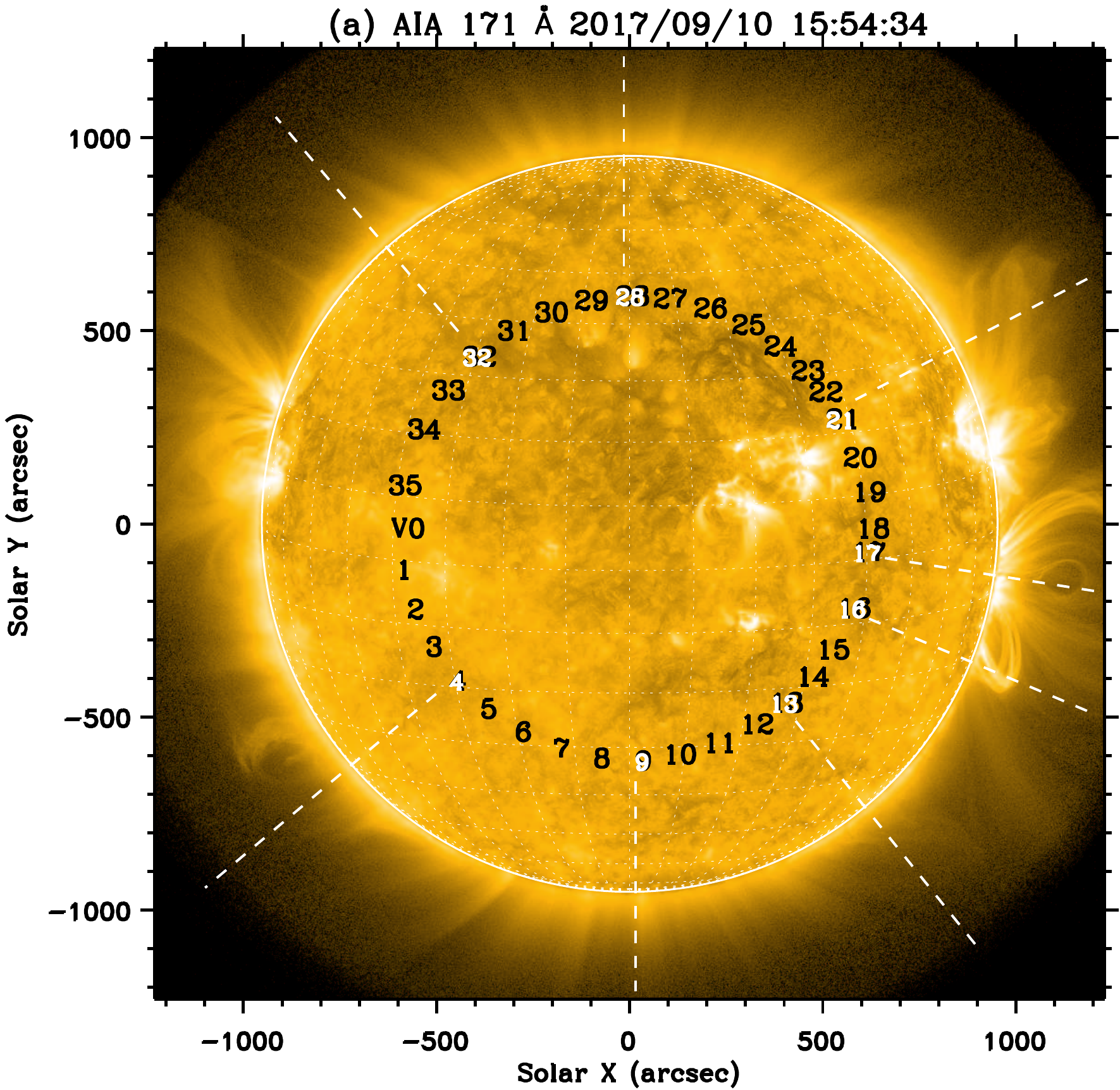} 
 \includegraphics[width=0.49\textwidth]{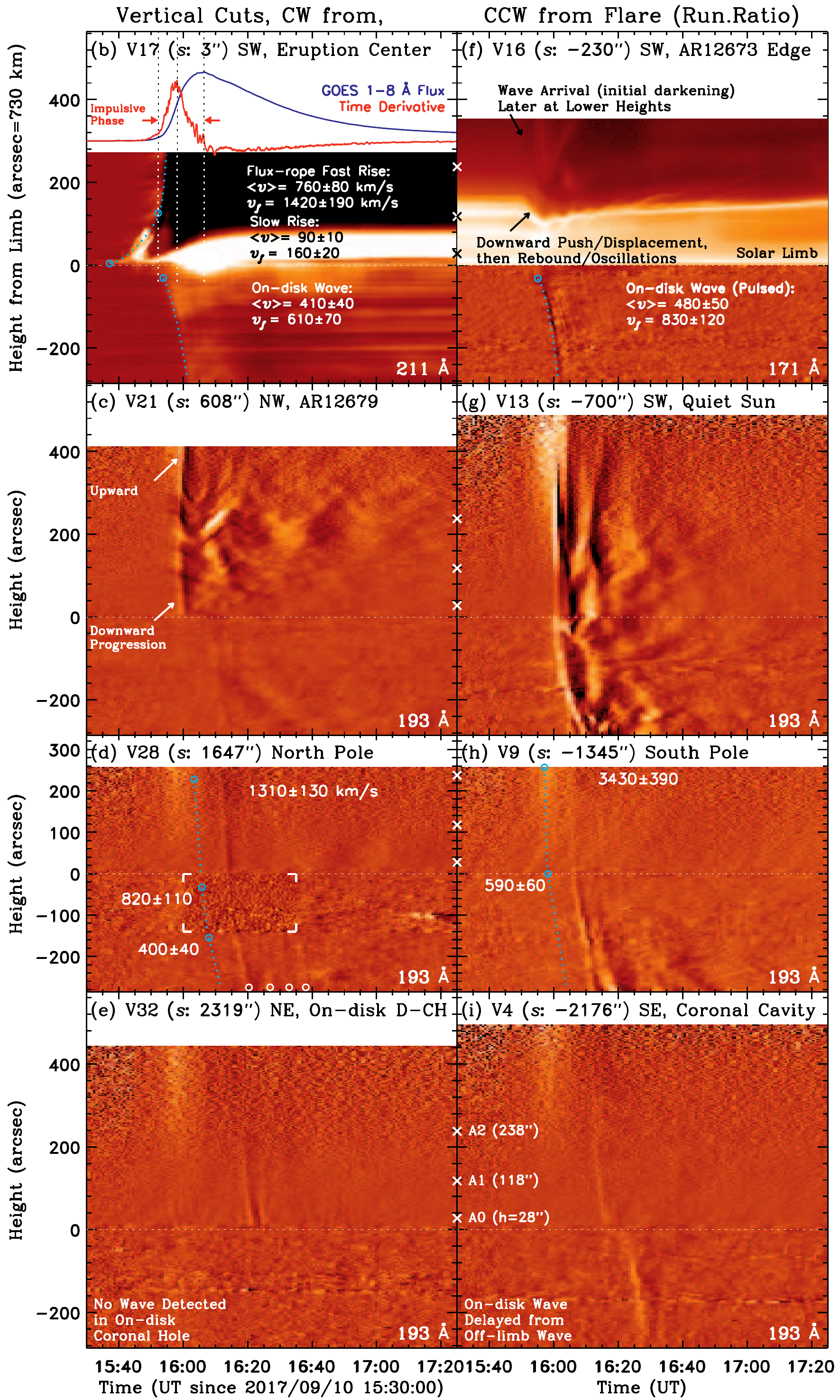} 
 \end{center}
 \vspace{-0.1in}
 \caption[]{
 Height dependence and off-limb to on-disk wave transition.
 (a) Context AIA 171~\AA\ image overlaid with selected vertical cuts in white.	
 (b)--(i) Corresponding \st\ plots in base-ratio at 211~\AA\ in (b) 
 and running-ratio 
 at 171/193~\AA\ in the other panels. The horizontal dotted line marks the limb position, 	
 above which the original 
 intensity is shown in (f). 
 The portion within the four white brackets (on-disk N-CH) here in (d) and those in \Figs{sector_slice.eps}(g)--(i)
 shows 171~\AA\ data to highlight the EUV wave.
 Cyan dotted lines in (b) and (f) 
 are parabolic fits (not shifted in time), labeled with average ($\langle v \rangle$) and final speeds ($v_f$).		
 } \label{vcut_slice.eps}
 \end{figure}

To examine height-dependent wave propagation,	
we placed 36 vertical cuts crossing the limb starting at $r= 0.7 R_\sun$. 	
\Fig{vcut_slice.eps} shows the resulting \st\ plots for 	
selected cuts. 	
Near the eruption (panel~(c)), 
the high-altitude wave front progresses upward due to the ascent and expansion of the CME bubble,
while the low-altitude front progresses downward (at large apparent speeds) due to 
the lateral expansion of the flux-rope from an elevated height,
causing a downward compression and the onset of the on-disk EUV wave (panel~(f)),
as previously revealed in Doppler observations 
\citep{Harra.Sterling.EIS.2010Feb16.EITwv.2011ApJ...737L...4H, Veronig.EIS.2010Feb16.EITwv.2011ApJ...743L..10V} 
and numerical simulations \citep[Figure~4(e) in][]{JinMeng.MHD.2011Feb15.CME.2016ApJ...820...16J}.
This, combined with the expected higher 	
fast-mode speeds at greater heights in the quiet-Sun corona, 
leads to the low-corona EUV wave front	
being forwardly inclined to the solar surface 
\citep{Uchida.Moreton-wave-sweeping-skirt.1968SoPh....4...30U, LiuW.cavity-oscil.2012ApJ...753...52L}.

This general pattern (negative slope in \st) holds not only in the off-limb low corona,
but also in the POS-projected on-disk portion within the vertical cuts. 	
This means that the on-disk wave closer to the limb 
arrives earlier, i.e., 
the near-limb wave mainly propagates along the limb,	
rather than across the disk. 	
There are occasional temporal discontinuities at the limb, 
e.g., 	
in the coronal cavity where the off-limb wave travels faster 	
\citep[also see Figure~5(e) in][]{LiuW.cavity-oscil.2012ApJ...753...52L}
than its on-disk counterpart 	
(\Figs{mosaic.eps}(g) and \ref{vcut_slice.eps}(i)).
%


\subsection{Waves in Polar Coronal Holes and Beyond}
\label{sect_CH}

One of this event's novel features	
is the strong transmissions 
\citep[cf.,][]{OlmedoOscar.2011Feb15_X2.AIA.EUV.wave.2012ApJ...756..143O}
into {\it both polar CHs}, besides 	
the commonly observed reflections 
\citep{Gopalswamy.EIT-wave-reflect-CH-illusion.2009ApJ...691L.123G}.
This, instead of apparent ``transmission"	
by line-of-sight (LOS) projection,	
is supported by the following evidence:

\begin{enumerate}

\item 
The EUV wave captured on the off-limb azimuthal cuts 
has elevated speeds (by $\times$[2\,--\,3]) within the CHs, 	
consistent with the expected higher fast-mode speeds.
If the wave were to propagate around (in front of or behind) the CHs, 
the POS-projected wave speed would not increase but remain the same or smaller.

\item
The vertical cuts within the polar CHs show clear continuation between the off-limb and on-disk waves,	
with different \st\ slopes but without detectable time lag at the limb (\Figs{vcut_slice.eps}(d) and (h)).
This suggests that the off-limb and on-disk waves are of the same front
that travels primarily along the limb. 	
The N-CH in particular extends substantially both off-limb and on-disk
(down to latitude $\sim$$60\degree$), throughout which the wave signal is present
and has a steeper slope than its quiet-Sun on-disk counterpart.

\item	
Another indication that the EUV wave actually travels into the polar CHs is the
wave-triggered transverse displacements (up to $60\kmps$) of polar plumes within CHs
that occur sequentially at increasing distances upon the wave arrival	
(\Fig{mosaic.eps}(a) animation)	
and appear as feather-like patterns in \st\ plots
(\Figs{limb_slice.eps}(h)--(i)). 

\end{enumerate}

An interesting effect of CH transmission is the alteration of
the wave propagation direction.
As shown in \Figs{mosaic.eps}(f) and (g),
the wave fronts emerging from the polar CHs are nearly parallel to latitudes,	
rather than longitudes	
(as one would expect for waves 	
from a source at the west limb), 
and travel toward the equator.
This is because the fast-mode speed increases toward the center of the CH,
causing the wave 	
to be refracted away from the center
and travel nearly radially outward. 
This effect, as numerically demonstrated	
\citep[][their Figure 4(e)]{Schmidt.Ofman.3D-MHD-eitwv.2010ApJ...713.1008S, JinMeng.MHD.2011Feb15.CME.2016ApJ...820...16J},
makes each polar CH essentially serve as a new ``radiation center" for the waves emerging from it. 

We find multiple wave fronts or ``ripples" leaving the polar CHs, at quasi-periodic intervals
in the 3--10~minute range, marked by white open circles in \Figs{vcut_slice.eps}--\ref{sector_slice.eps}.
As partly indicated in simulations 
\revise{\citep{Schmidt.Ofman.3D-MHD-eitwv.2010ApJ...713.1008S,
PiantschitschIsabell.EUV.wave.at.CH-3.wave.amplitude.2018ApJ...860...24P,
AfanasyevAndrei.Zhukov.EITwv.interact.CH.AR.2018AA...614A.139A}},
these pulses could result from a combination of:	
(i) direct reflection at the CH boundary in 3D,
(ii) refraction of the transmitted wave as noted above,
(iii) multiple reflections/bounces within the CH between its two end boundaries,
each producing its own transmission out of the CH,
and (iv) dispersive propagation of the primary fast-mode wave,
which itself exhibits certain periodicities near the eruption	
(e.g., \Figs{vcut_slice.eps}(f) and \ref{sector_slice.eps}(b)).
Pulsed waves from CHs were reported before \citep[e.g., Figure 6 in][]{YangLH.AIA.EIS.EUV-wave.2013ApJ...775...39Y},
but the large number of pulses (up to six) in this event is remarkable.	

\subsection{On-disk Wave Propagation}
\label{sect_on-disk}

\begin{figure*}[thbp]      
 \begin{center}
 \includegraphics[height=2.9in]{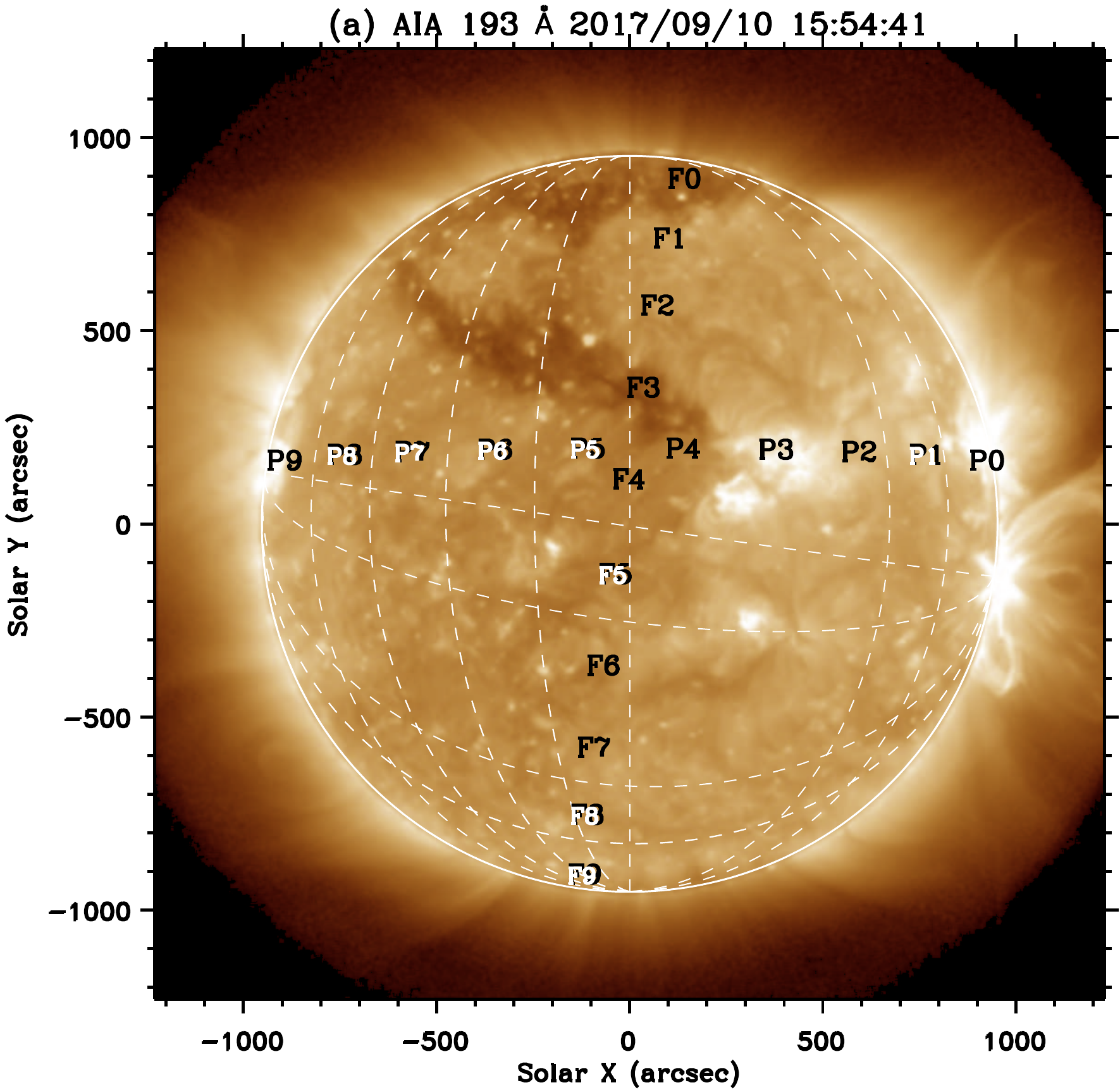} 
 \includegraphics[height=3.2in]{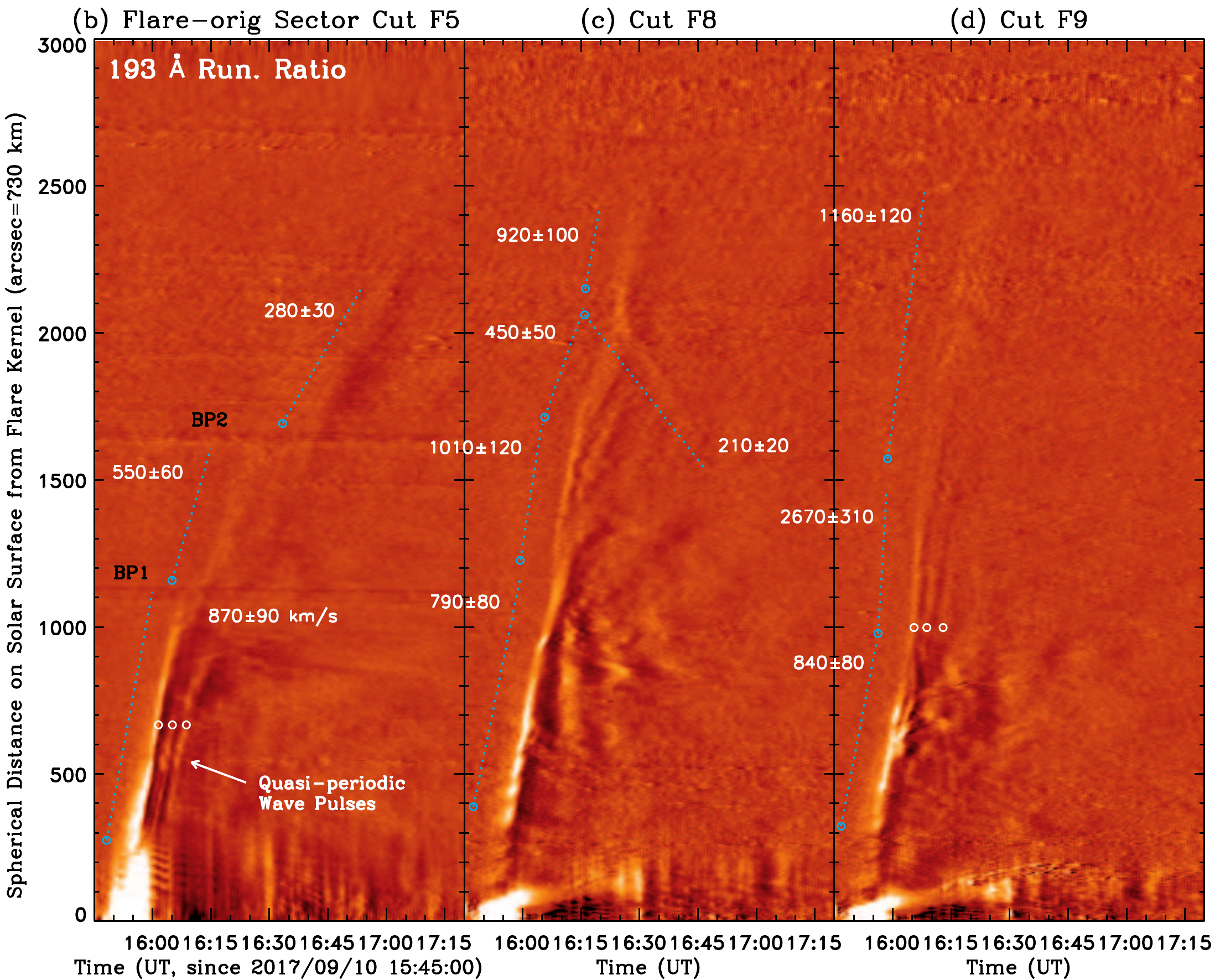} 
 \\
 \includegraphics[height=3.5in]{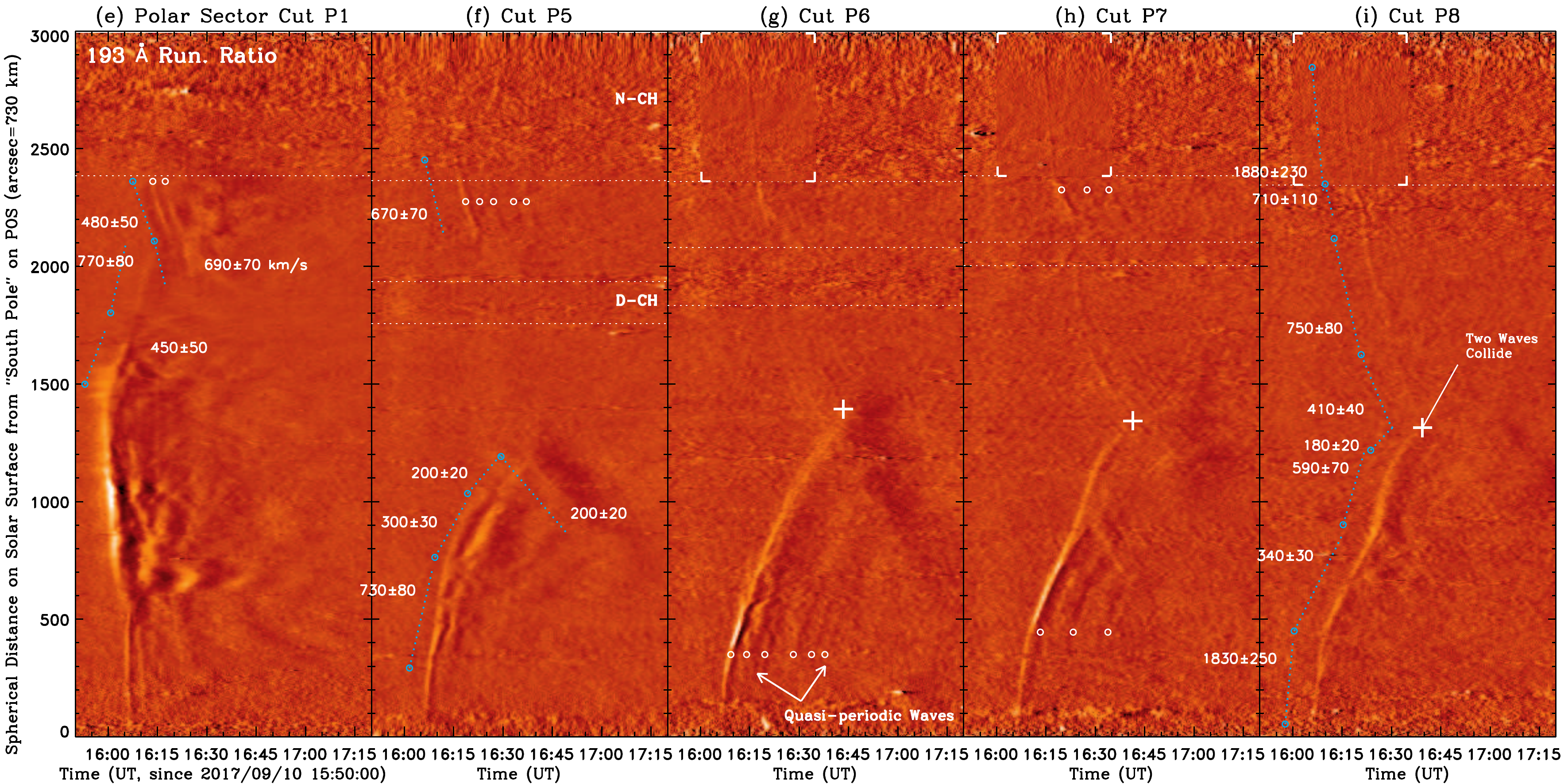} 
 \end{center}
 \vspace{-0.15in}
 \caption[]{
 On-disk wave propagation.
 (a) Context AIA 193~\AA\ image, overlaid with spherical sector cuts in white 	
 to obtain (b)--(i) \st\ plots. 
 The other sector cuts used are labeled in black, 	
 but not shown.
 The horizontal dotted lines in (e)--(i) mark the N-CH and D-CH boundaries.
 } \label{sector_slice.eps}
 \end{figure*}

To track the on-disk wave propagation 
and measure the so-call ``ground speed" 	
projected onto the	spherical solar surface	
\citep[e.g.,][]{LiuW.AIA-1st-EITwave.2010ApJ...723L..53L},
we employed two sets of spherical sector cuts: 	
one set (F0--F9) originating from the flare kernel at the limb
and the other (P0--P9) from the POS-projection of the Sun's south pole,
whose selected \st\ plots are shown in \Fig{sector_slice.eps} (top and bottom, respectively).

As shown in \Fig{mosaic.eps}, the initially circular-shaped on-disk wave front
is interrupted by a cluster of strong magnetic-field regions to the northeast/east of the eruption,	
including ARs and bipolar regions.
Only its southern portion	
advances substantially onto the quiet-Sun disk toward the east.	
This is also seen in \Figs{sector_slice.eps}(b)--(d),
where we find initial wave speeds of $\sim$$800\kmps$ near the eruption, 	
similar to the off-limb speeds (\Fig{limb_slice.eps}).
The speed generally decreases with distance, either gradually 
or abruptly	upon encountering 
local structures, 	
e.g., bipoles BP1 and BP2.	
The increased speeds in \Figs{sector_slice.eps}(c)--(d)	
are overestimates due to the wave from the S-CH approaching the cuts sideways. 	

The rest of the on-disk wave fronts \editor{comprise} 	
reflected and refracted secondary waves from the two polar CHs
that travel equatorward	
(\Figs{mosaic.eps}(f)--(h)) and are well captured by the polar sector cuts 
(\Fig{sector_slice.eps}, bottom).
They emerge from the poles at $\sim$$1800\kmps$, also comparable to 
the off-limb wave speeds, 	
and decelerate down to 200--$400\kmps$, within the expected range of quiet-Sun fast-mode speeds.
Eventually, the two equatorward waves 	
collide near BP2 around 16:40~UT, as marked by the plus signs, 
and produce extra intensity enhancements at 193/211~\AA\
followed by long-lasting, C-shaped dimming 	
(brightening at 171~\AA) that expands toward the southeast	
(\Fig{mosaic.eps}(i)).
Such interactions of counter-propagating waves 
\citep[e.g.,][]{OfmanLeon.LiuW.model.2013May22.counter.QFPs.2018ApJ...860...54O}
can \revise{result in} 
plasma heating, e.g., by turbulence generation and dissipation.


%

\subsection{Thermal Response of the Global Corona}
\label{sect_thermal}




As alluded earlier, 	
the EUV wave generally causes substantial intensity decreases in cool channels (171~\AA)	and	
increases in warm channels (193/211~\AA, with slight delays at 211~\AA). 	
This implies plasma heating across these channels' characteristic temperatures, from $\log T=5.9$ ($T=0.8\MK$)
to $6.2$ ($1.6\MK$) and then $6.3$ ($2.0\MK$).
This 	
is usually followed by an opposite change,	
suggestive of cooling. 
In composite running-ratio images (\Fig{mosaic.eps}), 
heating (cooling) corresponds to yellow/red (blue). 	
Such variations can be 
understood as wave-produced adiabatic compressional heating followed by rarefactional cooling
\citep{	
LiuW.cavity-oscil.2012ApJ...753...52L, 
DownsC.MHD.2010-06-13-AIA-wave.2012ApJ...750..134D}. 	

\begin{figure*}[thbp]      
 \begin{center}
 \includegraphics[height=3.5in]{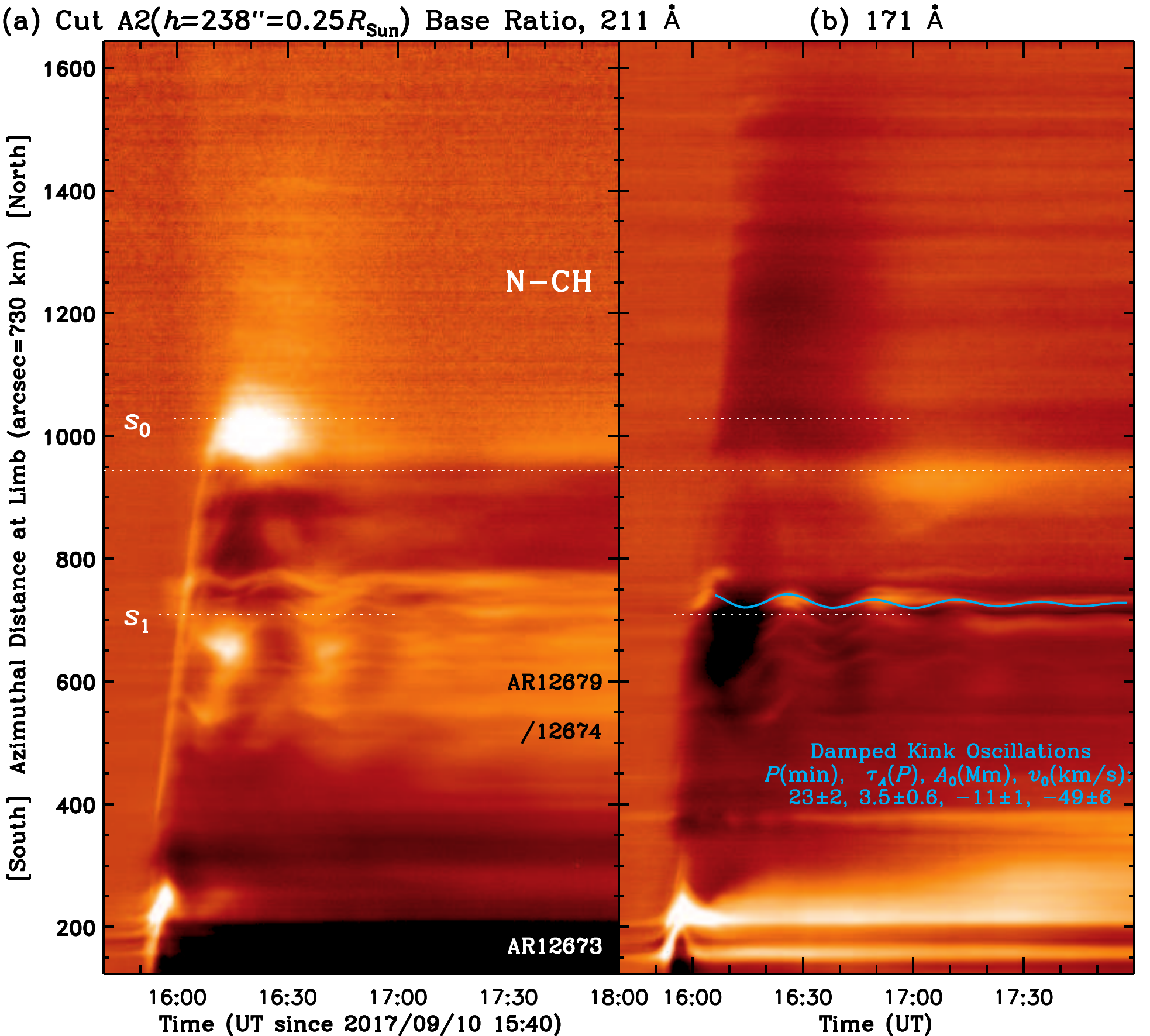} 
 \includegraphics[height=3.5in]{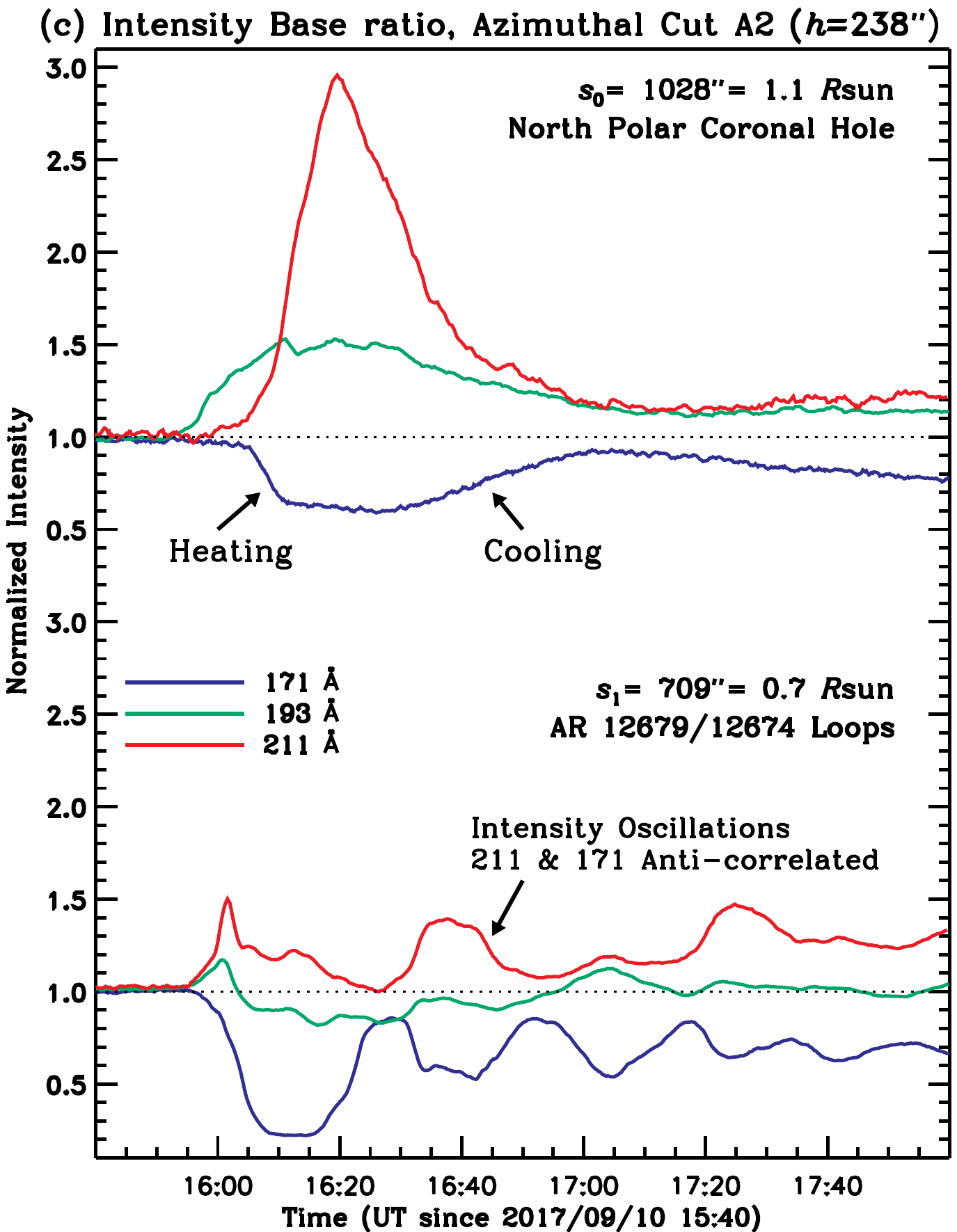} 
 \includegraphics[height=3.5in]{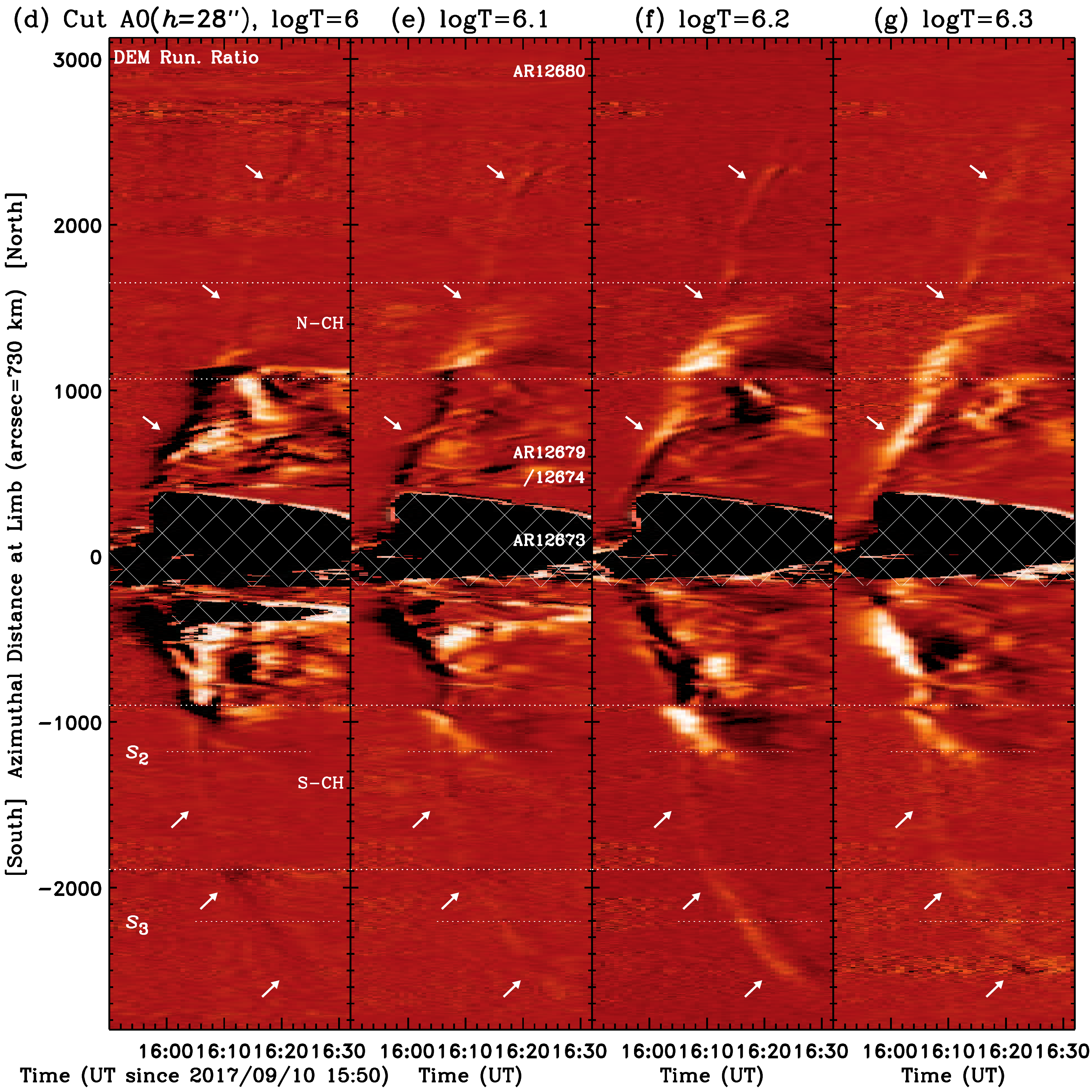} 
 \includegraphics[height=3.5in]{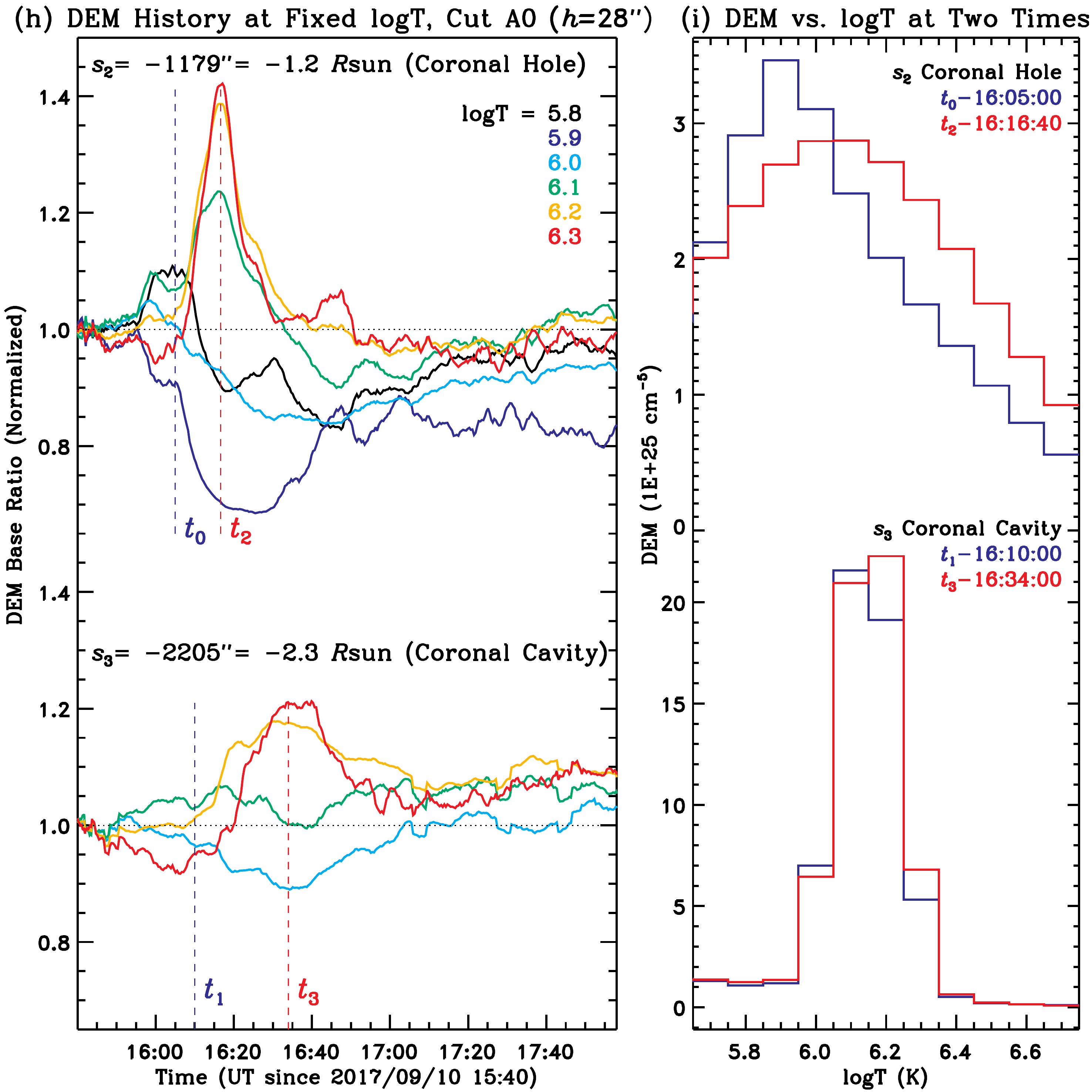} 
 \vspace{-0.15in}	
 \end{center}
 \caption[]{
 EUV wave's thermal impact	
 on the global corona. 
 Selected locations ($s_0$--$s_3$) are marked by short horizontal lines here on the left
 and by asterisks in \Fig{limb_slice.eps}(a).	
  (a)--(b) 211 and 171~\AA\ base-ratio \st\ plots from off-limb azimuthal Cut~A2 covering
 the Sun's northwest quadrant. 	
  (c) Corresponding intensity profiles at $s_0$ and $s_1$. 	
 (d)--(g) DEM running-ratio \st\ plots from 	
 Cut~A0 at four temperatures.	
 Hatched dark regions indicate failed DEM-inversion 	
 because of flare saturation. 	
  (h) Base-ratio DEM temporal profiles 	
 of selected temperatures at $s_2$ and $s_3$. 	
 (i) Corresponding DEM distributions, 	
 each at two times, before and after the wave impact and heating.	
 } \label{thermal.eps}
 \end{figure*}

\Figs{thermal.eps}(a)--(c) show an example of such anti-correlated intensity variations	
off the northwest limb.	
Within the N-CH at $s_0=1.1\Rsun$ from the eruption,	
the 211~\AA\ emission brightens by $\times3$	
from its pre-event level, while the 171~\AA\ intensity drops by $40\%$.
At $s_1=0.7\Rsun$	
in the northern peripheral of AR~12674/12679, such changes 
are $+50\%$ and $-80\%$, respectively.
These numbers	
dwarf those in mild EUV waves, e.g., $\lesssim$20\% in a C3.3 flare
\citep{LiuW.cavity-oscil.2012ApJ...753...52L} and $\lesssim$80\% in an M1.0 flare
\citep{DownsC.MHD.2010-06-13-AIA-wave.2012ApJ...750..134D}.
Such intensity variations can occur repeatedly involving multiple heating\,--\,cooling cycles,
e.g., at $500\arcsec<s<800\arcsec$ around $s_1$, 
which are associated with damped (by leakage and/or dissipation) kink oscillations of coronal loops, 	
$\Delta s(t)=A_0\exp(-t/\tau_A)\sin(2\pi t/P)$,
at a typical period $P=(23\pm2)$~minutes, damping time $\tau_A=(3.5\pm0.6)P$, 
and initial displacement and velocity amplitudes of $A_0=(11\pm1)\Mm$ and $v_0=(49\pm6)\kmps$.
%
%

To tease out 	
wave-caused subtle thermal changes,	
we performed differential emission measure (DEM) inversion \citep{CheungMark.AIA.DEM.invers.2015ApJ...807..143C},
which is generally underused 	
for EUV waves \citep{VanninathanK.DEM.EUV.wave.2011Feb15.2015ApJ...812..173V}.
We focused on the off-limb corona and constructed \st\ plots at selected temperatures from DEM maps.
As shown in \Figs{thermal.eps}(d)--(g) for azimuthal Cut~A0,
the EUV wave, marked by white arrows, is well captured.
Similar to the intensity variations noted above, 
the DEM generally decreases at lower temperatures (e.g., $\log T=6.0$)
and 	
increases at 	
higher temperatures (e.g., $\log T=6.2$),
indicating plasma heating, 	
followed by an opposite change, indicating recovery cooling.
The exact 	
change varies,	
depending on the distance from the eruption (thus wave or compression amplitude)
and the initial local DEM distribution.
For example, as shown in \Figs{thermal.eps}(h)--(i), 
at $s_2=-1.2\Rsun$		
inside the S-CH where the plasma is initially cool,	
the wave causes	
substantial, prompt (within $\sim$10~minutes) DEM increases at $\log T\ge6.1$ by $\lesssim$40\% 
and decreases at lower temperatures by $\lesssim$30\%, with the DEM peak shifted from  $\log T=5.9$ to $6.1$. 
Further away at $s_3=2.3\Rsun$	
in the quiet-Sun coronal cavity where the plasma is warm,		
the heating is gentle and gradual, with the DEM at $\log T\ge6.2$ increased
by $\lesssim$20\% and the peak temperature shifted from $\log T=6.1$ to $6.2$.

\section{Conclusion}	
\label{sect_conclude}



We have presented \sdo/AIA observations of an extraordinary global EUV wave associated with
the X8.2\editor{+} flare-CME on 2017 September 10. Major findings include \editor{the following.}

\begin{enumerate}	

\item	
This truly global	
EUV wave and a cascade of its secondary waves 	
were observed, for the first time, to traverse the {\it entire visible solar disk and off-limb corona}.

\item	
In addition to commonly observed reflections, 	
there are strong {\it transmissions into  both polar CHs}
at elevated speeds, which		
are then refracted out of the CHs	
toward the equator and eventually 
collide head-on. 	

\item	
The wave causes {\it large-amplitude thermal perturbations and structural oscillations},
some lasting for hours,	
signifying its profound impact on 	
the global coronal	
plasma. 	

\end{enumerate}		


These remarkable	
characteristics 	
opened a new window of utilizing EUV waves of such magnitudes
to probe the solar corona on global scales. 
This 	
allows {\it global coronal seismology}, an area yet to be fully exploited, to perform a variety of diagnostics 
to infer the physical conditions of the entire corona.
For example, 	
wave reflections and transmissions at polar CHs	
offer clues to	
the fast-mode speed and thus magnetic-field strength 	
in the polar regions, an important but poorly observed quantity.	
Using the measured EUV wave speeds and DEM-inferred density,
we obtained preliminary	
$B=9-12\G$ in the polar CHs and $3-6\G$ on the quiet-Sun
at the Cut~A1 height ($0.12\Rsun$).
Such analyses, together with numerical modeling providing direct comparison with observations, 
will be presented in future publications \editor{(M.~Jin et al., in preparation)}.


\editor{
\acknowledgments
{\sdo\ is the first mission of NASA's Living With a Star Program.
This work was supported by 	
NASA \sdo/AIA contract NNG04EA00C to LMSAL,
NASA grants NNX14AJ49G, NNX15AR15G, and NNX16AF78G, 
and NSF grant AGS-1259549. 
W.L. thanks Marc DeRosa for help with AIA tri-color movies and the anonymous referee for constructive comments.
}
}









{\scriptsize

}

\end{document}